\documentclass[pra,twocolumn,showpacs,superscriptaddress,amsmath]{revtex4-1}
\usepackage{graphicx}
\usepackage{upgreek}
\usepackage{color}
\usepackage{amsfonts}
\usepackage{gensymb}

\newcommand{\ket}[1]{\left|{#1}\right>}
\newcommand{\bra}[1]{\left<{#1}\right|}
\newcommand{\tr}[1]{\textnormal{tr}{\left\{#1\right\}}}
\newcommand{\I}{\mathrm{i}}
\newcommand{\E}{\mathrm{e}}
\newcommand{\D}{\mathrm{d}}

\begin{document}

\title{Incomplete quantum state estimation: a comprehensive study}

\author{Yong Siah Teo}
\affiliation{Centre for Quantum Technologies, National University of Singapore, Singapore 117543, Singapore}
\affiliation{NUS Graduate School for Integrative Sciences and Engineering, Singapore 117597, Singapore}
\author{Bohumil Stoklasa}
\affiliation{Department of Optics, Palacky University, 17. listopadu 12, 77146 Olomouc, Czech Republic}
\author{Berthold-Georg Englert}
\affiliation{Centre for Quantum Technologies, National University of Singapore, Singapore 117543, Singapore}
\affiliation{Department of Physics, National University of Singapore, Singapore 117542, Singapore}
\author{Jaroslav {\v R}eh{\'a}{\v c}ek}
\affiliation{Department of Optics, Palacky University, 17. listopadu 12, 77146 Olomouc, Czech Republic}
\author{Zden{\v e}k Hradil}
\affiliation{Department of Optics, Palacky University, 17. listopadu 12, 77146 Olomouc, Czech Republic}
\pacs{03.65.Ud, 03.65.Wj, 03.67.-a}

\begin{abstract}
We present a detailed account of quantum state estimation by joint maximization of the likelihood and the entropy. After establishing the algorithms for both perfect and imperfect measurements, we apply the procedure to data from simulated and actual experiments. We demonstrate that the realistic situation of incomplete data from imperfect measurements can be handled successfully.
\end{abstract}

\date{\today}

\begin{widetext}
\maketitle
\end{widetext}

\section{Introduction}
Quantum state preparation is the first important step for any protocol that makes use of quantum resources. Examples of such protocols are quantum state teleportation and quantum key distribution which require entangled quantum states. In order to verify the integrity of the quantum state of the source prepared, one carries out \emph{quantum state tomography} on the source. Measurements are performed on a collection of quantum systems (electrons, photons, etc.) that are emitted from the source, that is, a \emph{quorum}. Then, the quantum state of the source is inferred from the measurement data obtained from this ensemble. The measurements are generically described by a set of positive operators $\Pi_j$ that compose a \emph{probability operator measurement} (POM). The procedure of state inference, which shall be our main focus in this article, is also known as \emph{quantum state estimation}. If the size of the ensemble is infinite, the estimation procedure will yield the unique \emph{true} quantum state of the source; this is the frequentist's definition of the true state, which we accept as the best description of what the source prepares. However, such an ensemble is never achievable in any laboratory setting, as one can only perform measurements on a \emph{finite} ensemble of quantum systems. As a result, the state estimator obtained will be different from the true state and depends on the details of the estimation procedure. To make statistical predictions, the corresponding operator $\hat\rho$ describing this estimator must be a \emph{statistical operator}, which is positive. This will ensure that the estimated probability $\hat p_j=\tr{\hat\rho\Pi_j}$ for an outcome $\Pi_j$ of \emph{any} set of POM is positive. We shall denote all estimated quantities with a ``hat'' symbol.

There are two popular methods for quantum state estimation: \emph{Bayesian} and \emph{maximum-likelihood} (ML). The Bayesian state estimation method \cite{bayesian1,bayesian2,bayesian3} constructs a state estimator from an integral average over all possible quantum states. The \emph{likelihood functional}, which yields the likelihood of obtaining a particular sequence of measurement detection with a given quantum state, serves as a weight for the average. This approach includes all the neighboring states near the maximum of the likelihood functional as possible guesses for the unknown $\rho_\text{true}$. These neighboring states are given especially significant weight when $N$ is small, in which case the likelihood functional is only broadly peaked at the maximum. However, the integral average unavoidably depends on how one measures volumes in the state space, and there is no universal and unambiguous method for that. The ML approach \cite{ml1,ml2,ml3,ml4}, on the other hand, simply chooses the estimator as the statistical operator that maximizes the likelihood functional. Rather than identifying a unique estimator, as the Bayesian approach always does, the ML method may only yield a convex set of estimators if the estimated probabilities $\hat p_j$ are consistent with more than one statistical operator. If the ML estimator is unique, and the quorum sufficiently large, both approaches give the same estimator since the likelihood functional peaks very strongly at the maximum.

When the measurement outcomes form an \emph{informationally complete} set, the measurement data obtained will contain maximal information about the source. Thus, a unique state estimator can be inferred with ML. Unfortunately, in tomography experiments performed on complex quantum systems with many degrees of freedom, it is not possible to implement such an informationally complete set of measurement outcomes. As a result, some information about the source will be missing and its quantum state cannot be completely characterized. The ML estimator obtained from these informationally incomplete data is no longer unique and there will in general be infinitely many other ML estimators which are consistent with the data. In Ref.~\cite{main}, we briefly reported an iterative algorithm (MLME) to estimate unknown quantum states from incomplete measurement data by maximizing the likelihood and \emph{von Neumann entropy} functionals. In that Letter, we assumed that the measurement detections are perfect with no detection losses, i.e. $\sum_j\Pi_j=1$. The application of this algorithm was illustrated with examples of homodyne tomography and we concluded that, together with a more objective Hilbert space truncation, this approach can serve as a reliable and statistically meaningful quantum state estimation with incomplete data.

In this article, we will present more details on the recently proposed MLME algorithm and apply it to various other situations. First, we give a brief review of the mathematical formalism for quantum state estimation in Sec.~\ref{sec:formalism} to set the stage for the subsequent discussions. Next, we derive the numerical MLME algorithms respectively for \emph{both} perfect and imperfect measurement detections in Sec.~\ref{sec:algo}, with the latter being particularly useful for actual experiments. We illustrate applications of the two algorithms with two examples in Sec.~\ref{sec:app} and finally conclude in Sec.~\ref{sec:conc}.

\section{Formalism of quantum state estimation}
\label{sec:formalism}
In a tomography experiment, an ensemble of $N$ copies of quantum systems, identically prepared, is measured using a POM which consists of positive measurement outcomes $\Pi_j$. For simplicity, we first assume that all measurement detections are perfect and hence $\sum_j\Pi_j=1$. The problem of imperfect detections will be dealt with in Sec.~\ref{subsec:algo_imperfect}. For each outcome, its number of occurrences is denoted by $n_j$ such that $\sum_jn_j=N$. The likelihood functional $\mathcal{L}(\{n_j\};\rho)$, for a particular sequence of independent detections, is then
\begin{equation}
\mathcal{L}(\{n_j\};\rho)=\prod_jp_j^{n_j}\,.
\label{like1}
\end{equation}
As a consequence of perfect measurement detections, $\sum_jp_j=1$. The ML procedure searches for the estimator $\hat\rho_\text{ML}$ which maximizes $\mathcal{L}(\{n_j\};\rho)$. For a $D$-dimensional Hilbert space, when a POM comprises $D^2$ or more measurement outcomes, of which $D^2$ of them are linearly independent, it is informationally complete. In this case, there exists a unique estimator $\hat\rho_\text{ML}$ for a given set of measurement data $\{n_j\}$. One can also define the outcome frequencies $f_j=n_j/N$ out of these measurement data such that $\sum_jf_j=1$. The corresponding functional $\mathcal{L}(\{n_j\};\rho)$ due to this informationally complete POM will peak at the unique global maximum $\hat\rho_\text{ML}$ over the space of $\rho$, whereby $\hat\rho_\text{ML}$ is solely determined by the frequencies $f_j$ and does not depend on the total number $N$ of measured copies.

The situation is different when the POM is informationally incomplete. In this case, there will be infinitely many ML estimators satisfying a smaller set of linearly independent constraints imposed by the incomplete measurement data. These ML estimators form a convex set of operators which maximize the convex functional $\mathcal{L}(\{n_j\};\rho)$. Geometrically, $\mathcal{L}(\{n_j\};\rho)$ possesses a convex plateau structure hovering over the space of $\rho$. The task, now, is to select one of these estimators for future statistical predictions. To do this, we adopt the well-known maximum-entropy (ME) principle advocated by Jaynes \cite{jaynes}. That is, we look for the estimator with the largest von Neumann entropy
\begin{equation}
S(\rho)=-\tr{\rho\log\rho}\,
\label{ent}
\end{equation}
among the convex set of ML estimators. This supplementary step introduces a small and smooth convex hill over the plateau structure so that a unique maximum can be obtained. The corresponding MLME estimator $\hat\rho_\text{MLME}$ is the least-bias estimator for the given set of incomplete measurement data; it can be regarded as the most conservative guess of the unknown quantum state out of the convex set of ML estimators.

At this point, we would like to comment on the distinction between this MLME technique and the conventional ME technique \cite{me1,me2}. The ME technique takes the outcome frequencies $f_j$ as \emph{bona fide} estimates for the probabilities $p_j$ and tries to search for the positive operator
\begin{equation}
\hat\rho_\text{ME}=\frac{\E^{\sum_j\lambda_j\Pi_j}}{\tr{\E^{\sum_j\lambda_j\Pi_j}}}
\end{equation}
that maximizes $S(\rho)$, subjected to the probability constraints which are mediated by the Lagrange multipliers $\lambda_j$. The fundamental problem with this scheme is that the $f_j$s cannot always be treated as probabilities since there may not be \emph{any} statistical operator $\rho$ for which $f_j=\tr{\rho\Pi_j}$. This is due to the statistical noise which is inherent in the outcome frequencies arising from measuring a finite ensemble of quantum systems. Therefore, in such cases, the ME technique fails as there simply is no positive operator which is consistent with the measurement data to begin with. The MLME algorithm, on the other hand, looks for the unique MLME estimator by confining the search within the plateau region inside the space of statistical operators. Thus, positivity is ensured. In cases where the $f_j$s are probabilities, both the ME and MLME schemes yield the same estimator by construction since the estimated probabilities $\hat p_j=f_j$ correspond to a statistical operator.

\section{The numerical algorithms}
\label{sec:algo}
\subsection{Perfect measurements}
\label{subsec:algo_perfect}
Assuming that the measurement detections are perfect, the likelihood functional $\mathcal{L}(\{n_j\};\rho)$ in Eq.~(\ref{like1}) gives a complete statistical description of all possible sequences of detections for the $N$ measured copies of quantum systems. Equivalently, one can consider the optimization of the \emph{normalized log-likelihood functional} $\log(\mathcal{L}(\{n_j\};\rho))/N$ to simplify the subsequent calculations, in view of the monotonic nature of the logarithmic function. The motivation for introducing the normalization will become clear soon. The MLME scheme can then be perceived as a standard constrained optimization problem: maximize $\log(\mathcal{L}(\{n_j\};\rho))/N$ subjected to the constraint that $S(\rho)$ takes the maximal value $S_\text{max}$. This is equivalent to maximizing $S(\rho)$ with the constraint that $\log(\mathcal{L}(\{n_j\};\rho))/N$ is maximal, as discussed above. The Lagrange functional for this optimization problem is defined as
\begin{equation}
\mathcal{I}(\lambda;\rho)=\lambda \bigl(S(\rho)-S_\text{max}\bigr)+\frac{1}{N}\log\mathcal{L}(\{n_j\};\rho)\,,
\end{equation}
where $\lambda$ is the Lagrange multiplier corresponding to the constraint for $S(\rho)$. We denote the estimator that maximizes $\mathcal{I}(\lambda;\rho)$ by $\hat\rho_{\text{I},\lambda}$. Incidently, the functional $\mathcal{I}(\lambda;\rho)$ is a sum of two different types of entropy, up to an irrelevant additive constant $\sum_jf_j\log f_j$: the von Neumann entropy $S(\rho)$ that quantifies the ``lack of information'', and the \emph{negative} of the \emph{relative entropy} $S(\{f_j\}|\{p_j\})=\sum_jf_j\log(f_j/p_j)$ that quantifies the ``gain of information'' from the measurement data. The scheme can now be interpreted as a simultaneous optimization of two complementary aspects of information, with an appropriately assigned constant relative weight $\lambda$. In addition, the normalization of $\log\mathcal{L}(\{n_j\};\rho)$ renders the optimal value of $\lambda$ to be independent of $N$.

\begin{figure}
\centering
\includegraphics[width=0.5\textwidth]{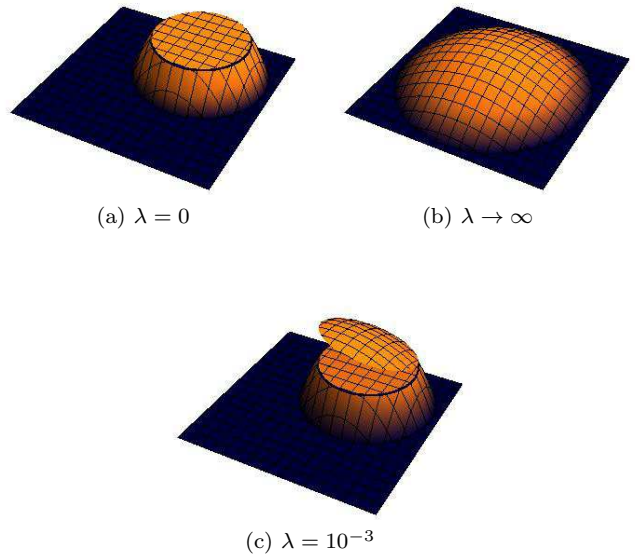}
  \caption{Schematic diagrams of $\mathcal{I}(\lambda,\rho)$ on the space of statistical operators. The maximally-mixed state resides at the center of the square base which represents the Hilbert space. At the extremal points of $\lambda$, $\mathcal{I}(\lambda=0;\rho)=\log(\mathcal{L}(\{n_j\};\rho))/N$, with a convex plateau at the maximal value, and $\mathcal{I}(\lambda\rightarrow\infty;\rho)=\lambda S(\rho)$. Plot (c) shows the functional with an appropriate choice of value for $\lambda$ for MLME. An additional hill-like structure resulting from $S(\rho)$ is introduced over the plateau, so that the estimator with the largest entropy can be selected from the convex set of ML estimators within the plateau.}
  \label{fig:geom}
\end{figure}

When $\lambda=0$, we recover the Lagrange functional for the log-likelihood functional alone. Owing to the informational incompleteness of the measurement data, there exists a convex plateau structure for the log-likelihood functional. As $\lambda\rightarrow\infty$, the von Neumann entropy becomes increasingly more significant and the resulting estimator $\hat\rho_{\text{I},\lambda\rightarrow\infty}$ approaches the maximally-mixed state $1/D$. Naturally, when $\lambda$ takes on a very small positive value, the contribution from $\lambda S(\rho)$ becomes much smaller than $\log(\mathcal{L}(\{n_j\};\rho))/N$ and the effect of the von Neumann entropy functional is only significant over the plateau region in which the likelihood is maximal. Figure~\ref{fig:geom} illustrates all the aforementioned points.
This means that, in general, $\lambda$ should be chosen so small that $S\left(\hat\rho_{\text{I},\lambda}\right)$ is very close to the minimum, and below which there are only very slight changes in the two entropy functionals \cite{main}.

Let us derive the iterative algorithm for maximizing $\mathcal{I}(\lambda\rightarrow 0;\rho)$ with respect to $\rho$. After varying $\mathcal{I}(\lambda\rightarrow 0;\rho)$, we have
\begin{equation}
\updelta\mathcal{I}(\lambda\rightarrow 0;\rho)=-\lambda\,\tr{\updelta\rho\log\rho}+\sum_j\frac{f_j}{p_j}\updelta p_j\,.
\label{var1}
\end{equation}
The variations $\updelta p_j$, or $\updelta\rho$, have to be such that $\rho$ stays positive after these variations. To choose their appropriate forms, we first parameterize the positive operator $\rho=\mathcal{A}^\dagger \mathcal{A}/\tr{\mathcal{A}^\dagger \mathcal{A}}$ with an auxiliary complex operator $\mathcal{A}$. Under this parametrization,
\begin{equation}
\updelta\rho=\frac{\updelta \mathcal{A}^\dagger \mathcal{A}+\mathcal{A}^\dagger\updelta \mathcal{A}-\rho\,\tr{\updelta \mathcal{A}^\dagger \mathcal{A}+\mathcal{A}^\dagger\updelta \mathcal{A}}}{\tr{\mathcal{A}^\dagger \mathcal{A}}}\,.
\label{varrho}
\end{equation}
Substituting $\updelta\rho$ in Eq.~(\ref{varrho}) into Eq.~(\ref{var1}), we have
\begin{equation}
\updelta\mathcal{I}(\lambda\rightarrow 0;\rho)=\tr{\frac{\updelta \mathcal{A}^\dagger \mathcal{A}}{\tr{\mathcal{A}^\dagger \mathcal{A}}}\mathfrak{R}+\mathfrak{R}\frac{\mathcal{A}^\dagger \updelta \mathcal{A}}{\tr{\mathcal{A}^\dagger \mathcal{A}}}}\,,
\label{var2}
\end{equation}
where
\begin{equation}
\mathfrak{R}=R-1-\lambda\bigl(\log\rho-\tr{\rho\log\rho}\bigr)
\end{equation}
with
\begin{equation}
R=\sum_j\frac{f_j}{p_j}\Pi_j\,.
\end{equation}

When $\mathcal{I}(\lambda\rightarrow 0;\rho)$ is maximal, we have $\updelta\mathcal{I}(\lambda\rightarrow 0;\rho)=0$ and the extremal equations
\begin{equation}
\rho\,\mathfrak{R}=\mathfrak{R}\rho=0
\label{exteqn1}
\end{equation}
are satisfied. Therefore, to solve these extremal equations numerically, we iterate the equation
\begin{equation}
\rho_\text{k+1}=\frac{\left(\mathcal{A}^\dagger_k+\updelta \mathcal{A}^\dagger_k\right)\left(\mathcal{A}_k+\updelta \mathcal{A}_k\right)}{\tr{\left(\mathcal{A}^\dagger_k+\updelta \mathcal{A}^\dagger_k\right)\left(\mathcal{A}_k+\updelta \mathcal{A}_k\right)}}\,
\label{itereqn1}
\end{equation}
starting from some statistical operator $\rho_1$, until $k=k'$ such that the norm of $\rho_{k'} \mathfrak{R}_{k'}$ is less than some pre-chosen value. We then take $\hat\rho_\text{MLME}\equiv\rho_{k'}$ as the MLME estimator. Maximizing $\mathcal{I}(\lambda\rightarrow 0;\rho)$ will require $\updelta\mathcal{I}(\lambda\rightarrow 0;\rho)$ to be positive whenever $\mathcal{I}(\lambda\rightarrow 0;\rho)$ is less than the maximal value. A straightforward way to enforce positivity is to set
\begin{equation}
\updelta \mathcal{A}_k\equiv\left(\updelta \mathcal{A}^\dagger_k\right)^\dagger\equiv\epsilon \mathcal{A}_k\mathfrak{R}_k\propto\epsilon\frac{\partial \mathcal{I}(\lambda;\rho)}{\partial \mathcal{A}_k}\,,
\label{itereqn2}
\end{equation}
with $\epsilon$ being a small positive constant. This is the \emph{steepest-ascent} method. We have thus established a numerical MLME scheme as a set of iterative equations (\ref{itereqn1}) and (\ref{itereqn2}) to search for the MLME estimator using the measurement data obtained from perfect measurement detections. More compactly, the relevant iterative equations are
\begin{align}
  \rho_\text{k+1}&=\frac{\left(1+\epsilon \mathfrak{R}_k\right)\rho_k\left(1+\epsilon \mathfrak{R}_k\right)}{\tr{\left(1+\epsilon \mathfrak{R}_k\right)\rho_k\left(1+\epsilon \mathfrak{R}_k\right)}}\,,\nonumber\\
  \mathfrak{R}_k&=R_k-1-\lambda\left(\log\rho_k-\tr{\rho_k\log\rho_k}\right)\,.
  \label{mlmeiteralgo1}
\end{align}
We note that a more efficient algorithm, using the conjugate-gradient method, can be derived from this steepest-ascent algorithm, which is the subject of a separate discussion.

\subsection{Imperfect measurements}
\label{subsec:algo_imperfect}
In actual experiments, the measurement detections will usually be imperfect in the sense that the detection efficiency $\eta_j$ of a particular measurement outcome $\Pi_j$ is less than unity. In this case, the overall outcome probabilities
\begin{equation}
\tilde p_j\equiv\eta_jp_j
\end{equation}
will not sum to unity. Hence, we have a set of POM with outcomes $\tilde\Pi_j\equiv\eta_j\Pi_j$ such that $G\equiv\sum_j\tilde\Pi_j<1$. A consequence of this is that the true total number $M$ of copies received is not known, since only $N<M$ are detected ($N=M$ when all $\eta_j=1$ as in Sec.~\ref{subsec:algo_perfect}).

The likelihood functional that accounts for all $M$ copies of quantum systems in an experiment with imperfect detections is given by
\begin{equation}
\tilde{\mathcal{L}}(\{n_j\};\rho)=\frac{M!}{N!\,(M-N)!}\left(\prod_j\tilde p_j^{n_j}\right)\left(1-\eta\right)^{M-N}\,,
\label{like2}
\end{equation}
where $\eta=\sum_j\tilde p_j<1$. The additional combinatorial prefactor arises from the indistinguishability in the ordering of the detection sequence resulted from losses. With the help of Stirling's approximation for the factorials, the variation of the corresponding log-likelihood functional is given by
\begin{align*}
\updelta\log\tilde{\mathcal{L}}(\{n_j\};\rho)&=\tr{\left(N\tilde R-\frac{M-N}{1-\eta}G\right)\updelta\rho}\nonumber\\
&+\updelta M\log\left(\frac{(1-\eta)M}{M-N}\right)\,,
\end{align*}
where $\tilde R=\sum_jf_j\tilde\Pi_j/\tilde p_j$. Adopting the concept of maximum-likelihood, we derive an expression for $M$ such that $\log\tilde{\mathcal{L}}(\{n_j\};\rho)$ is maximized for any given $\rho$. This implies that the coefficient of the arbitrary $\updelta M$ must vanish and we have $M=N/\eta$ as the most-likely value of $M$. With this, the expression for $\tilde{\mathcal{L}}(\{n_j\};\rho)$ reduces to the simple form
\begin{equation}
\tilde{\mathcal{L}}(\{n_j\};\rho)=\prod_j\left(\frac{p_j}{\eta}\right)^{n_j}
\label{rel_like}
\end{equation}
up to an irrelevant multiplicative factor, with its corresponding logarithmic variation
\begin{equation}
\updelta\log\tilde{\mathcal{L}}(\{n_j\};\rho)=N\tr{\left(\tilde R-\frac{G}{\eta}\right)\updelta\rho}\,.
\label{varlike2}
\end{equation}
The additional term $-\updelta\rho G/\eta$ in the argument of the trace accounts for copies that have escaped detection.

Defining $\mathcal{I}(\lambda\rightarrow 0;\rho)$ for the new POM and its $\tilde{\mathcal{L}}(\{n_j\};\rho)$ in Eq.~(\ref{rel_like}), one can derive the iterative equations
\begin{align}
\rho_{k+1}&=\frac{\left(1+\epsilon \tilde{\mathfrak{R}}_k\right)\rho_k\left(1+\epsilon\tilde {\mathfrak{R}}_k\right)}{\tr{\left(1+\epsilon\tilde {\mathfrak{R}}_k\right)\rho_k\left(1+\epsilon \tilde {\mathfrak{R}}_k\right)}}\,,\nonumber\\
\tilde {\mathfrak{R}}_k&=\tilde R_k-\frac{G}{\eta^{(k)}}-\lambda\left(\log\rho_k-\tr{\rho_k\log\rho_k}\right)\,,
\label{iteralgo2}
\end{align}
with $\eta^{(k)}=\sum_j\tilde p^{(k)}_j$.

\begin{figure}[h!]
  \centering
  \includegraphics[width=0.4\textwidth]{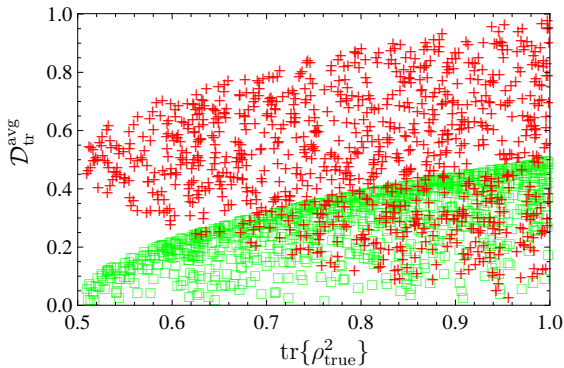}
  \caption{A comparison of two different schemes with $10^3$ random qubit true states distributed uniformly with respect to the Hilbert-Schmidt measure. Fifty experiments were simulated for every true state, with $N=5000$ for each experiment, and the respective average trace-class distances $\mathcal{D}^\text{avg}_\text{tr}$ were computed. The entire simulation was done with a set of randomly generated, informationally incomplete POM consisting of two imperfect measurement outcomes. The plot markers denoted by ``$+$'' represent reconstructed states using the algorithm in Eq.~(\ref{mlmeiteralgo1}) while ignoring the imperfection of the measurements, and those denoted by ``$\square$'' represent the reconstructed states using the algorithm in Eq.~(\ref{iteralgo2}) that accounts for this imperfection. The significant improvement in tomographic efficiency with the latter algorithm is a strong indication of the importance of a proper treatment of imperfect measurements.}
  \label{fig:imp_det}
\end{figure}

To highlight the importance of a proper treatment of imperfect measurement detections, we perform a simulation on $10^{3}$ randomly generated qubit states. Figure~\ref{fig:imp_det} compares the performance of the MLME algorithm derived in Sec.~\ref{subsec:algo_perfect}, with which we search for the MLME estimator by assuming that the measured data $\{n_j\}$ are all we have while ignoring the possible missing data, with that of the MLME algorithm derived in this section. The trace-class distance
\begin{equation}
\mathcal{D}_\text{tr}=\frac{1}{2}\tr{|\hat\rho_\text{MLME}-\rho_\text{true}|}
\end{equation}
is used as the figure of merit to quantify the distance between $\hat\rho_\text{MLME}$ and $\rho_\text{true}$. The lesson here is that if one neglects the consequence of imperfect measurements in performing state reconstruction, the quality of the resulting reconstructed state estimator will typically be much lower than that obtained from a scheme which accounts for this imperfection.

\section{Applications}
\label{sec:app}
\subsection{Time-multiplexed detection tomography}
\label{subsec:tmd}
\begin{figure}
\centering
\includegraphics[width=0.4\textwidth]{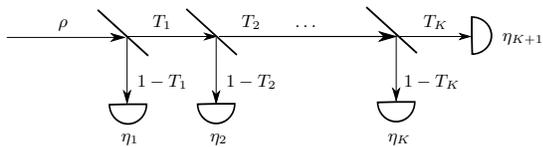}
  \caption{A schematic diagram representing the time-multiplexed setup with $K+1$ output ports. The $T_j$s are the respective transmission probabilities for the $j$th beam splitter. The overall efficiency for, say, the $k$th port is given by $\tilde{\eta}_k=\eta_k(1-T_k+T_{K+1}\delta_{k,K+1})\prod^{k-1}_{j=1} T_j$.}
  \label{fig:tmd_diag}
\end{figure}

First, we apply the MLME technique to simulation experiments on \emph{time-multiplexed detection} (TMD) tomography \cite{TMD}. For experiments of this type, photon pulses, of a particular quantum state, containing more than one photon are sent through a series of beam splitters \cite{remark1}, each associated with a certain transmission probability. Behind each of the output ports of such a series is a single-photon detector that either registers a click from an incoming split photon pulse, with some detection efficiency, or does nothing. Thus, each output port has a certain overall efficiency $\tilde{\eta}_j$ which is related to the relevant transmission probabilities and detection efficiency (See Fig.~\ref{fig:tmd_diag}).

As a consequence of this, the POM outcomes
\begin{equation}
\Pi_{j}=\sum_n\ket{n}c_{jn}\bra{n}
\end{equation}
will be a mixture of Fock states, with the coefficients $c_{jn}$ related to $\eta_j$ \cite{fiberloop}. If there are $N_\text{ports}$ output ports, where \emph{all} $\eta_j$s are different, there will be $2^{N_\text{ports}}$ distinct POM outcomes due to the binary nature of the single-photon detectors. In addition, $\sum^{2^{N_\text{ports}}}_{j=1}\Pi_j=1$ since the $2^{N_\text{ports}}$ binary sequences of detection configurations constitute all possible events. These POM outcomes commute and a measurement of these outcomes only gives information about the diagonal entries of the statistical operator of the true state in the Fock basis. In order to obtain information about the off-diagonal entries, one can, for instance, displace the current set of $2^{N_\text{ports}}$ POM outcomes in phase space with some complex value $\alpha_k$ away from the origin using the displacement operator
\begin{equation}
\mathcal{D}(\alpha_k)=\E^{\alpha_k A^\dagger-\alpha^*_k A}\,,
\end{equation}
where $A$ is the standard photon annihilation operator. Then, the new set of outcomes
\begin{equation}
\Pi_j(\alpha_k)=\frac{1}{\mathcal{N}}\mathcal{D}(\alpha_k)\Pi_j\mathcal{D}^\dagger(\alpha_k)\,,
\end{equation}
with $\mathcal{N}$ being the total number of such displaced set of $2^{N_\text{ports}}$ outcomes, do not commute with the undisplaced set. These displaced outcomes are suitable for a measurement that is designed to obtain information about the unknown true state by sampling over multiple $\alpha_k$s. Experimentally, these displaced POM outcomes can be realized with unbalanced homodyne detection \cite{unbalanced}.

In the simulations, four output ports, corresponding to a total of $2^4=16$ POM outcomes, are considered. Two different true states are selected to illustrate the results of MLME. The first true state is chosen to be a stationary state of a laser given by
\begin{equation}
\rho_{\text{ss}}=\E^{-\mu}\sum^{\infty}_{n=0}\ket{n}\frac{\mu^n}{n!}\bra{n}
\label{laser_ss}
\end{equation}
where $\mu$ is the mean number of photons \cite{laserss}. For the second true state, the statistical operator \mbox{$\rho_{\alpha'}=\ket{\textsc{m}(\alpha')}\bra{\textsc{m}(\alpha')}$}, where
\begin{equation}
\ket{\textsc{m}(\alpha')}=\frac{\ket{\alpha'}+\ket{-\alpha'}}{\sqrt{2\left(1+\E^{-2|\alpha'|^2}\right)}}\,
\label{schro_cat}
\end{equation}
is the superposition of the coherent states $\ket{\alpha'}$ and $\ket{-\alpha'}$, is chosen. The notation $\ket{\textsc{m}(\alpha')}$ is used to denote the ket for the \emph{male} ``Schr{\"o}dinger's cat'' state. See, for example, Ref.~\cite{catref} for a survey of the family of cat states. Statistical operators are first reconstructed from the simulated data. For this reconstruction, one has to decide on the dimension $D_\text{sub}$ of the truncated Hilbert space for the reconstructions. This procedure, also commonly known as \emph{state-space truncation}, depends on the prior information about the unknown state. In our case, suppose one knows that the mean number of photons of the source is $\mu\approx 4$, which is the value assigned in the simulation. Then, one may anticipate that all the relevant information about the true state should be contained in a Hilbert space of a dimension which is close to $\mu$. In fact, it is a common practice to choose $D_\text{sub}$, compatible with this information, such that the displaced operators form an informationally complete POM. Then, the standard ML method can be applied to state estimation. We shall compare the result of this approach with another, perhaps more objective, methodology in which we select a larger subspace compatible with this prior information and estimate the state with MLME.

After obtaining the reconstructed statistical operators, the Wigner functions $W(x,p)$ of the dimensionless position and momentum quadrature values, $x$ and $p$ respectively, are calculated in accordance with
\begin{align}
&W(x,p)\nonumber=2\E^{-|\alpha|^2}\sum^{\infty}_{m=0}\sum^{\infty}_{n=0}\bra{m}\rho\ket{n}\nonumber\\
\times &\left[(-1)^{j_<}\sqrt{\frac{2^{j_>}j_<!}{2^{j_<}j_>!}}(x+\I^{\,\text{sgn}(n-m)} p)^{|m-n|}L_{j_<}^{(|m-n|)}\left(2\,|\alpha|^2\right)\right]\,,
\label{wigner}
\end{align}
where $\alpha=x+\I p$ and $L_{n}^{(\nu)}(y)$ is the degree-$n$ \emph{associated Laguerre polynomial} in $y$ of order $\nu$, for all the statistical operators. Here, we define $j_<\equiv \min\{m,n\}$ and $j_>\equiv \max\{m,n\}$.

\begin{figure}
\centering
\includegraphics[width=0.5\textwidth]{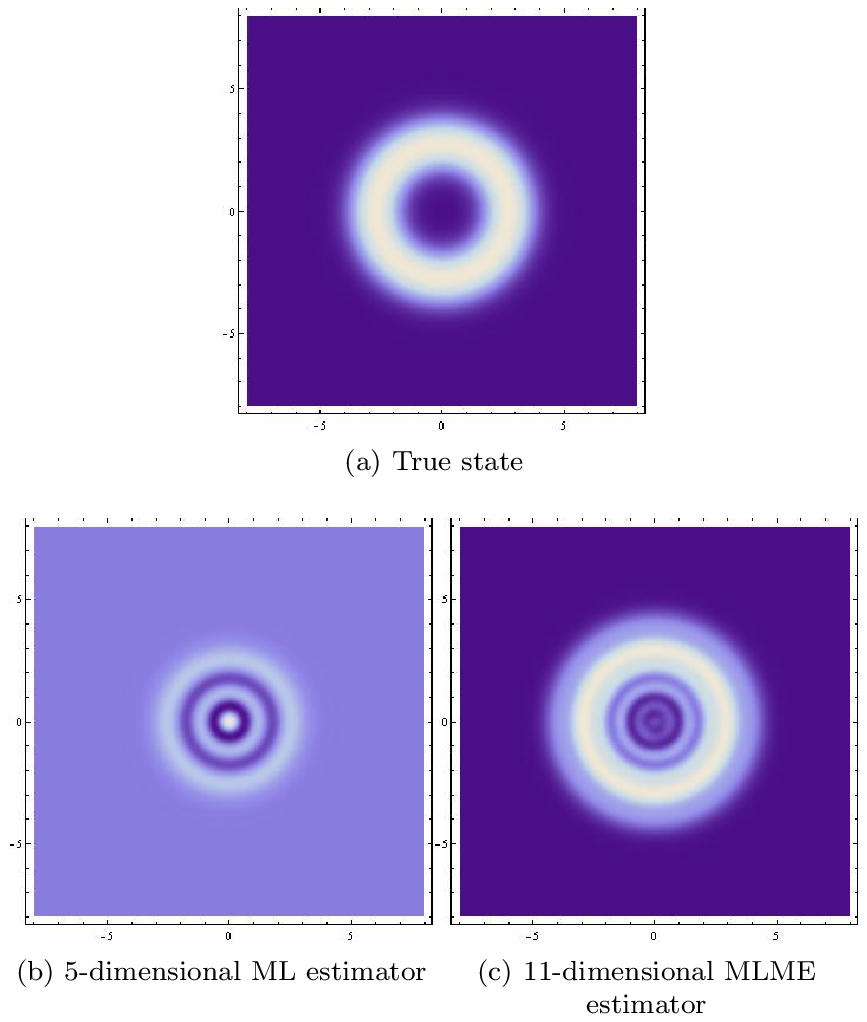}
  \caption{Density plots of the Wigner functions, in phase space, of various statistical operators for (a) the true state (20-dimensional stationary state of a laser, $\mu=4$) with $\tilde\tau\approx 0.394$, (b) the 5-dimensional ML estimator with $\tilde\tau\approx 0.921$ and (c) the 11-dimensional MLME estimator with $\tilde\tau\approx 0.489$. Here, brighter regions indicate the locations of larger Wigner function values, and vice versa. The statistical operator for (b) is obtained using ML by assuming a 5-dimensional subspace in which the displaced POM outcomes are informationally complete. The statistical operator for (c) is obtained by assuming a larger subspace of dimension 11 using MLME. Numerous artificial nonclassical features of the ML estimator, a signature of its highly oscillatory Wigner function, are manifested as an abnormally large value of $\tilde\tau$, an inevitable byproduct of state-space truncation. One can see that with MLME, extraneous artifacts of the Wigner function resulted from such a truncation can be largely removed.}
  \label{fig:tmd_density}
\end{figure}
\begin{figure}
\centering
\includegraphics[width=0.5\textwidth]{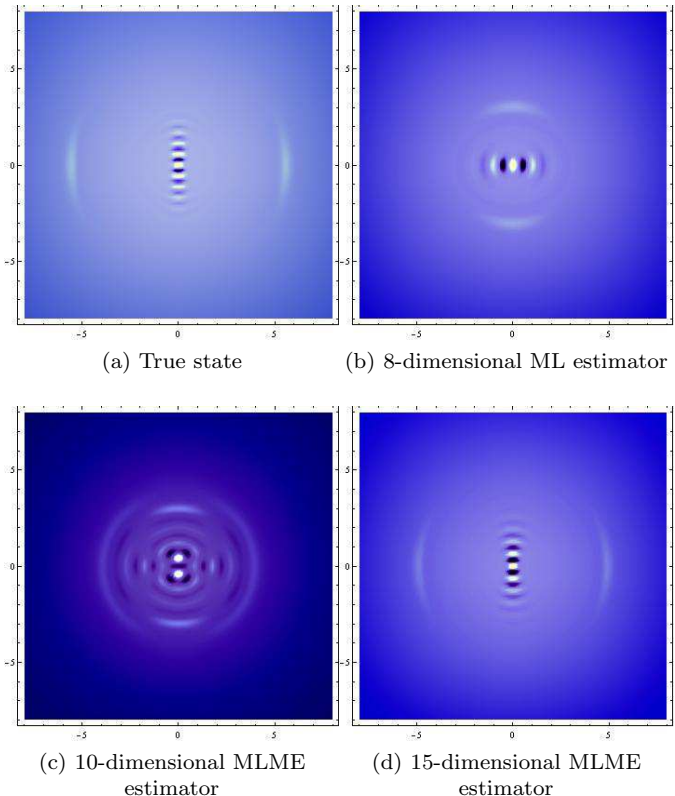}
  \caption{Density plots of the Wigner functions, in phase space, of various statistical operators for (a) the true state ($\rho_{\alpha'}$, $\alpha'=5$), (b) the 8-dimensional ML estimator, (c) the 10-dimensional and (d) 15-dimensional MLME estimators. In this case, the Wigner function of the ML estimator differs greatly from that of the true state, an example of misleading information obtained via state-space truncation. A transition in the structure of the Wigner function occurs at $D_\text{sub}=10$, with the MLME estimator for $D_\text{sub}=15$ giving a more accurate estimated picture of the Wigner function of the true state.}
  \label{fig:tmd_density_cat}
\end{figure}

To quantify the nonclassicality of the statistical operators, we make use of the concept of \emph{nonclassicality depth} introduced in Ref.~\cite{nonclassicalitydepth}. Let us define the function
\begin{equation}
\mathcal{R}(\alpha,\tau)=\frac{1}{\pi\tau}\int (\D w)\,\text{exp}\left(-\frac{|\alpha/\sqrt{2}-w|^2}{\tau}\right)P(w)\,,
\end{equation}
where $w$ is a complex variable, $(\D w)$ denotes the integral measure over the real and imaginary parts of $w$, $P(w)$ is the \emph{Glauber-Sudarshan $P$ function}, and the parameter $\tau$ is in the range $0\leq\tau\leq1$. From the above definition, it follows that $\mathcal{R}(\alpha,\tau)$ is a continuous interpolating function of $\tau$ from the typically singular, as well as non-positive, $P(\alpha/\sqrt{2})$ ($\tau\rightarrow 0$), to the Wigner function $W(\alpha)$ ($\tau=1/2$), and finally to the positive \emph{Husimi $Q$ function} $Q(\alpha)=\bra{\alpha}\rho\ket{\alpha}/2\pi$ ($\tau\rightarrow 1$).

The nonclassicality depth is then defined as the smallest value $\tau=\tilde\tau$, above which $\mathcal{R}(\alpha,\tau)\geq0$. Any mixture of coherent states is therefore a classical state since, in this case, $\tilde\tau=0$. A quantum state with $\tilde\tau>0$ is a nonclassical state. This measure of nonclassicality captures the nonclassical nature of quantum states through a one-parameter family of functions, which can otherwise be invisible to measures involving a fixed value of $\tau$, such as the conventional negativity of the Wigner function. Although quantifying nonclassicality with $\tilde\tau$ is a somewhat arbitrary procedure, we adopt it here as a measure of nonclassicality that is not worse than other proposals.

The generalization of \eqref{wigner} to arbitrary $\tau$ values,
\begin{align}
&\mathcal{R}(x,p,\tau)=\frac{\E^{-\frac{|\alpha|^2}{2\tau}}}{\tau}\sum^{\infty}_{m=0}\sum^{\infty}_{n=0}\bra{m}\rho\ket{n}\nonumber\\
\times \Bigg[&\,(-1)^{j_<}\sqrt{\frac{j_<!}{j_>!}}\left(\frac{1-\tau}{\tau}\right)^{j_>}\nonumber\\ &\,\left(\frac{x+\I^{\,\text{sgn}(n-m)} p}{\sqrt{2}(1-\tau)}\right)^{|m-n|}L_{j_<}^{(|m-n|)}\left(\frac{|\alpha|^2}{2\tau(1-\tau)}\right)\Bigg]\,,
\label{nonclass_eq}
\end{align}
is useful for the numerical computation of $\tilde\tau$. For the stationary state in Eq.~(\ref{laser_ss}), Eq.~(\ref{nonclass_eq}) simplifies to
\begin{align}
&\mathcal{R}_{\text{ss}}(x,p,\tau)\nonumber\\
=&\,\frac{\E^{-\frac{|\alpha|^2}{2\tau}-\mu}}{\tau}\sum^\infty_{n=0}(-1)^n\,\frac{\mu^n}{n!}\left(\frac{1-\tau}{\tau}\right)^nL_n\left(\frac{|\alpha|^2}{2\tau(1-\tau)}\right)\,.
\end{align}

The performances of both MLME and the standard ML method on the true states defined in Eqs.~(\ref{laser_ss}) and (\ref{schro_cat}) are illustrated by the Wigner function plots of the respective statistical operators obtained from both methods. These are shown in Figs.~\ref{fig:tmd_density} and \ref{fig:tmd_density_cat}. The respective nonclassicality depths are also computed for Fig.~\ref{fig:tmd_density}. For the state $\rho_{\alpha'}$, all the corresponding reconstructed statistical operators are highly nonclassical, with $\tilde\tau=1$ \cite{telenonclass} for all of them. Hence, rather than compare the $\tilde\tau$ values, the structure of the Wigner functions for various reconstruction subspaces will be briefly analyzed instead in Fig.~\ref{fig:tmd_density_cat}.

\subsection{Light-beam tomography}
\label{subsec:beam}

\begin{figure}
\includegraphics[width=0.8\columnwidth]{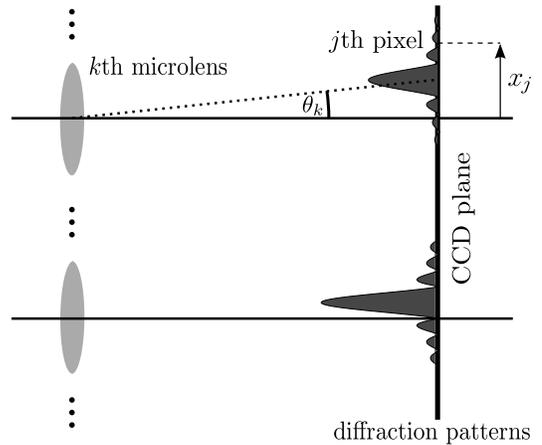}
\caption{Schematic diagram of the diffraction patterns of an incoming light beam that is obtained from a SH wave front sensor. The light beam is transformed by an array of microlenses (apertures). A CCD camera is placed at the rear focal plane of the array. The measurement data consist of the measured intensities of the beam. The intensity at the $j$th pixel, located at position $x_j$, behind the $k$th microlens aperture is denoted by $I_k(x_j)$.\label{fig:sh}}
\end{figure}

Finally, we make use of the MLME algorithm to reconstruct states of classical light beams that are measured using the Shack-Hartmann (SH) wave front sensor. An incoming light beam is transformed by a regular array of microlens apertures and detected in its rear focal plane by a charge-coupled device (CCD) camera (see Fig.~\ref{fig:sh}). A plane wave traversing in the transverse plane of the SH sensor gives rise to a detection, where the individual diffraction patterns are centered at the corresponding optical centers of the microlenses. For a distorted wave front, the observed diffraction pattern behind the $k$th microlens aperture will be deflected by an angle $\theta_k$. Since the set of angles $\theta_k$ is related to the local wave front tilts with respect to the transverse plane of the SH sensor, the shape of the wave front can be inferred. Clearly, this standard technique of wave front reconstruction fails in the presence of imperfect coherence, where the notions of ``wave front'' and ``optical phase'' are no longer well-defined and a more general description of the state of the light beam is necessary.

Recently, an alternative theory for SH detection, based on the principles of quantum state tomography, has been introduced. It was shown that a complete characterization of a beam of light is possible
from the measurement data obtained with the SH sensor under certain assumptions with regard to the aperture profiles \cite{coherence}. Analogously to quantum states, we can describe a coherent beam (mode), with a complex amplitude $\psi(x)$, by a ket $\ket{\psi}$, such that $\psi(x)=\langle x|\psi\rangle$. It should be understood, that this $\psi(x)$ is not a quantum mechanical probability amplitude, but a mathematical symbol with analogous properties that we exploit. At the focal plane of the $k$th microlens aperture, the amplitude $\psi'_k(x)$ of the transformed beam is given by
\begin{equation}
\label{transform}
\psi'_k(x)=\int\D x'\,h_k(x-x')a_k(x')\psi(x'),
\end{equation}
where $a_k(x)$ is the aperture function of the $k$th microlens aperture and the response function $h_k(x)$ describes the free propagation from the $k$th microlens to the SH sensor.

Now, suppose a generic \emph{partially} coherent beam is detected by the SH sensor. We can describe the state of such a beam with a \emph{coherence operator} $\rho_\text{coh}$. When using a computational basis of orthonormal modes $|\psi_n\rangle$, we have
\begin{equation}
\rho_\text{coh}\,=\,\sum_{mn}\ket{\psi_m}\rho^\text{coh}_{mn}\bra{\psi_n}.
\end{equation}
By defining the aperture operator
\begin{equation}
M^{(a)}_k=\int\D x'\,\ket{x'}a_k(x')\bra{x'}
\end{equation}
for the $k$th microlens aperture and the unitary propagation operator $U_k$, where $\bra{x}U_k\ket{x'}=h_k(x-x')$, that describes the free propagation from the $k$th microlens to the SH sensor, the representation of the corresponding transformed state $\rho'_\text{coh}$,
\begin{align}
\rho'_\text{coh}&=\,U_k\,M^{(a)}_k\,\rho_\text{coh}\,M^{(a)}_k\,U_k^\dagger\nonumber\\
&=\sum_{mn}\underbrace{U_k\,M^{(a)}_k\ket{\psi_m}}_{\equiv\ket{\psi'_m}}\rho^\text{coh}_{mn}\underbrace{\bra{\psi_n}M^{(a)}_k\,U_k^\dagger}_{\equiv\bra{\psi'_n}}\nonumber\\
&=\sum_{mn}\ket{\psi'_m}\rho^\text{coh}_{mn}\bra{\psi'_n}\,,
\end{align}
on the focal plane of the apertures follows from the linearity of optics transformations. The intensity $I_k(x_j)$ at position $x_j$ \cite{remark2} on the rear focal plane of the $k$th aperture is
\begin{equation}
\begin{split}
\label{intensity1}
I_k(x_j)&\equiv\langle x_j|\rho'_\text{coh}|x_j\rangle\\
&=\langle x_j| \bigg(\sum_{mn}\ket{\psi'_{m,j}}\rho^\text{coh}_{mn}\bra{\psi'_{n,k}}\bigg)
|x_j\rangle\\
&=\sum_{mn}\rho^\text{coh}_{mn}\,\psi'_{m,k}(x_j)\psi'_{n,k}(x_j)^*\,,
\end{split}
\end{equation}
where $\psi'_{n,k}(x_j)=\langle x_j|\psi'_{n,k}\rangle$ are the complex amplitudes of the transformed light beam obtained from the amplitudes $\psi_n(x_j)=\langle x_j|\psi_n\rangle$ of Eq.~\eqref{transform}. Since $\rho_\text{coh}$ possesses all the properties of a statistical operator, the MLME technique can be used to estimate the true coherence operator $\rho^\text{true}_\text{coh}$ of a partially coherent beam. To this end, we need to compute the corresponding POM describing the measurement outcomes of the SH sensor. By relating $I_k(x_j)$ to the corresponding probabilities of the outcomes $\Pi_k(x_j)\widehat{=}\sum_{mn}\ket{\psi'_m}\Pi_{k,nm}(x_j)\bra{\psi'_n}$, we have
\begin{equation}
\label{intensity2}
\begin{split}
I_k(x_j)&=\tr{\rho_\text{coh}\,\Pi_k(x_j)}\\
&=\sum_{mn}\rho^\text{coh}_{mn}\,\Pi_{k,nm}(x_j)\,.
\end{split}
\end{equation}
Comparing Eqs.~\eqref{intensity1} and \eqref{intensity2}, the positive operator describing the detection outcome at the
$j$th pixel of the CCD camera behind the $k$th aperture is given by
\begin{equation}
\label{POMelements}
\Pi_{k,nm}(x_j)=\psi'_{m,k}(x_j)\psi'_{n,k}(x_j)^*.
\end{equation}
As an illustrative example, the POM outcomes considered in this section are commuting operators in the infinite-dimensional Hilbert space with regard to the coherence operators. Equivalently, the aperture functions for the respective microlenses do not overlap in position. This is a \emph{special case} of a more general theory on Shack Hartmann detection, which will be discussed at length in another upcoming article.

\begin{figure}
\setlength{\unitlength}{0.8cm}
\begin{picture}(6,5)
\put(-4.5,-3){\includegraphics[width=1.35\columnwidth]{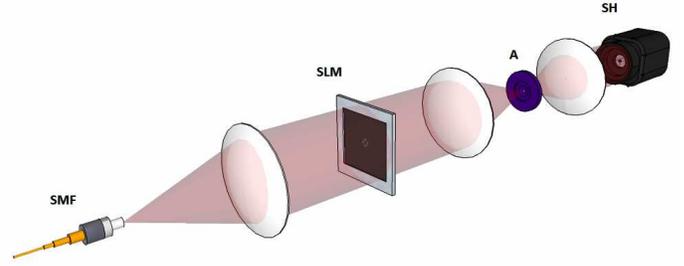}}
\end{picture}
\caption{Experimental set-up involving a single-mode fiber (SMF), a spatial light modulator (SLM), an aperture stop (A) and a Shack-Hartmann (SH) sensor. \label{fig:setup}}
\end{figure}

In the experiment, a controlled preparation of optical beams is realized using the principles of digital holography \cite{setup}. Figure~\ref{fig:setup} shows the set-up. The essence of the beam preparation lies in the numerical construction of a digital hologram that is programmed to produce a superposition of a reference plane wave and a beam with the true state $\rho^\text{true}_\text{coh}$ of interest. This is achieved with the help of an amplitude spatial light modulator (OPTO SLM) with a resolution of 1024$\times$768 pixels. The hologram is then illuminated by the reference plane wave that is considered in the superposition. To approximately produce this plane wave, a collimated Gaussian beam is generated by placing the output of a single-mode fiber at the focal plane of a collimating lens. In this way, the digital hologram can be fully situated at the center of the collimated Gaussian beam of a larger beam waist, where this beam can then be approximated to be a plane wave with high accuracy. The resulting diffraction spectrum, after illuminating the digital hologram with the collimated Gaussian beam, involves several diffraction orders, of which only one contains useful information about $\rho^\text{true}_\text{coh}$. To filter out the unwanted diffraction orders, a 4-$f$ optical processor, with a small circular aperture stop placed at the rear focal plane of the second lens, is used for this purpose (the aperture stop in Fig.~\ref{fig:setup}). The resulting light beam with the state $\rho^\text{true}_\text{coh}$ is then focussed at the rear focal plane of the third lens. This completes the preparation stage.

The measurement of the light beam involves a Flexible Optical SH sensor with 128 microlenses that form a hexagonal array. Each microlens has a focal length of 17.9mm and a hexagonal aperture with a diameter of 0.3mm. The signal at the focal plane of the array is detected by a uEye CCD camera that has a resolution of 640$\times$480 pixels, with each pixel being 9.9$\upmu$m$\times$9.9$\upmu$m in size.

The aforementioned set-up is used for generating and analyzing low-order Laguerre-Gaussian (LG) modes. The LG modes can serve as important resources in quantum information processing \cite{Lgmodes}. In this experiment, only LG modes with no radial nodes are considered. Such modes form a one-parameter orthonormal basis, where the modes are specified by the orbital angular momentum quantum number $l$. In polar coordinates, the relevant part of the complex amplitude of a LG mode, for a fixed $l$, is given by
\begin{equation}
\langle s,\varphi|\text{LG}_l\rangle\propto s^l \E^{\I l \varphi} \E^{-s^2}\,.
\end{equation}
Nonzero values of $l$ give rise to helical wave fronts, for which each photon carries an orbital angular momentum of $l\hbar$.

For the source of light beams, we would like to prepare the state $\rho^\text{true}_\text{coh}=\rho^\text{sup}_\text{coh}=\ket{\psi_\text{sup}}\bra{\psi_\text{sup}}$, where
\begin{equation}
\label{statetrue}
\ket{\psi_\text{sup}}=\left(\ket{\text{LG}_0}-\ket{\text{LG}_1}\I-\ket{\text{LG}_2}\right)\frac{1}{\sqrt{3}}\,,
\end{equation}
using the OPTO SLM. In the presence of experimental imperfections, however, the true state $\rho^\text{true}_\text{coh}$ prepared this way will not be exactly the same as $\rho^\text{sup}_\text{coh}$. After measuring this beam with the SH sensor, the data are processed using the MLME algorithm in Eq.~(\ref{iteralgo2}) to obtain the estimator $\hat\rho^\text{MLME}_\text{coh}$ for $\rho^\text{true}_\text{coh}$, since $G<1$. To quantify the quality of $\hat\rho^\text{MLME}_\text{coh}$, we investigate the \emph{fidelity} between $\hat\rho^\text{MLME}_\text{coh}$ and $\rho^\text{sup}_\text{coh}$.

\begin{figure}
\includegraphics[width=0.9\columnwidth]{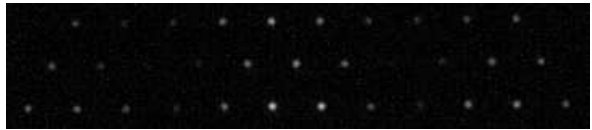}
\caption{CCD image for the state $\rho^\text{true}_\text{coh}$. The relevant part of the SH readout used for the beam reconstruction is shown. Contributions from the individual SH apertures are indicated by bright spots, with each spot made up of multiple pixels. Note that the two void regions correspond to the phase singularities of the state $\rho^\text{sup}_\text{coh}$. This hints that $\rho^\text{true}_\text{coh}\approx\rho^\text{sup}_\text{coh}$.
\label{fig:data}}
\end{figure}
Figure~\ref{fig:data} shows the CCD image for the state $\rho^\text{true}_\text{coh}$. Each aperture gives rise to a bright spot in the CCD image. To maximize the signal-to-noise ratio, only the pixel with the highest intensity within each spot is selected as a measurement datum. The set of intensities, corresponding to maximum-intensity pixels, constitute the measurement data to be used for state reconstruction. In our case, the corresponding POM consists of $35$ linearly independent outcomes described by Eq.~\eqref{POMelements}. This measurement is, therefore, informationally complete for $D_\text{sub}\leq5$.
\begin{figure}
\includegraphics[width=\columnwidth]{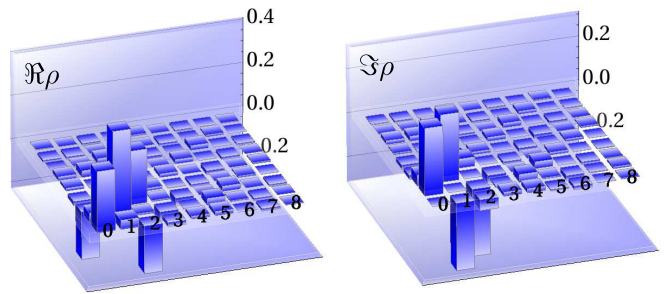}
\caption{MLME state estimation from informationally incomplete data for $D_\text{sub}=9$. The real (left) and imaginary (right) parts of the reconstructed coherence operator $\hat\rho^\text{MLME}_\text{coh}$ are shown. The reconstruction subspace is spanned by the modes $\text{LG}_l$, with $l=0,1,\ldots,8$. In this case, $56$ out of $91$ independent outcomes, required for complete characterization of $\rho^\text{true}_\text{coh}$, are not accessible, yet the MLME estimator $\hat\rho^\text{MLME}_\text{coh}$ is close to $\rho^\text{sup}_\text{coh}$, with a fidelity of $92\%$.
\label{fig:recon}}
\end{figure}
In cases where state reconstruction on informationally complete subspaces gives unsatisfactory results, the MLME approach can be used on the informationally incomplete data to give reasonable estimators on a larger subspace, as illustrated in Fig.~\ref{fig:recon}.

\begin{figure}[h!]
\vspace{10pts}
\includegraphics[angle=-90,width=0.75\columnwidth]{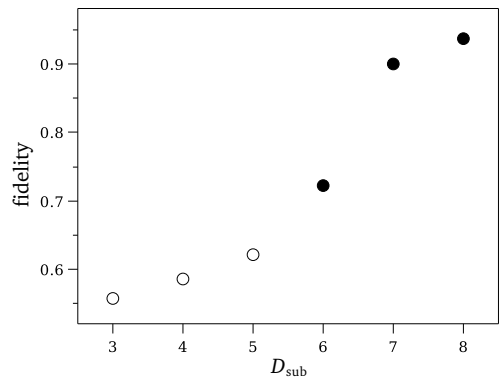}
\caption{Average fidelities, computed over 50 random choices of computational bases, of the estimators for different dimensions $D_\text{sub}$ of the reconstruction subspace. The unfilled (filled) circular plot markers correspond to informationally complete (incomplete) tomography,
respectively.
\label{fig:meanfid}}
\end{figure}

So far, the procedure of state-space truncation is performed in the basis of the $\text{LG}_l$ modes. In this basis, when $\rho^\text{true}_\text{coh}$ is known to be quite close to $\rho^\text{sup}_\text{coh}$, the truncation of modes of higher orders will not result in a great loss of reconstruction information, as implied by the structure of $\rho^\text{sup}_\text{coh}$ in Eq.~(\ref{statetrue}). The situation will be very different when there is no such prior knowledge about $\rho^\text{true}_\text{coh}$, except for the fact that the possible values of $l$ lie in a certain range. In this situation, there is no appropriate strategy to choose a computational basis in which the state-space truncation can be done effectively and justifiably. More generally, estimating the unknown state $\rho^\text{true}_\text{coh}$ on a truncated subspace will, as a rule, result in missing important reconstruction information and this will lead to strongly biased estimators. A remedy for this problem is to perform state reconstruction on a sufficiently large subspace that is compatible with the knowledge about the range of values of $l$.

To emphasize this point, we simulate the following scenario:
\begin{itemize}
\item The set of measurement data, obtained from the CCD image shown in Fig.~\ref{fig:data}, is distributed to $50$ parties. The possible values of $l$ for the true state $\rho^\text{true}_\text{coh}$ are known to lie in the range $l\in[0,7]$.
\item Each party selects a computational basis and estimates the state of the beam for $D_\text{sub}=3,4,\ldots,8$ using either the ML (for $D_\text{sub}\leq5$) or the MLME algorithm (for $D_\text{sub}>5$).
\item The reconstructed estimators for the six values of $D_\text{sub}$ are reported by each party and the average fidelity of the estimators for every value of $D_\text{sub}$ are calculated.
\end{itemize}
A typical outcome of this scenario is shown in Fig.~\ref{fig:meanfid}. As can be seen, performing state-space truncations in order to reconstruct $\rho^\text{true}_\text{coh}$ with an informationally complete set of data generally leads to low fidelities in the estimators. Increasing the number of degrees of freedom and using the MLME algorithm to cope with the completeness issue seems to be a much better strategy.

\section{Conclusion}
\label{sec:conc}

We derived the iterative algorithms for informationally incomplete quantum state estimation respectively for perfect and imperfect measurements. Next, we applied these algorithms to time-multiplexed detection tomography and light-beam tomography. From these two applications, we learned that one should better not restrict the state reconstruction to a subspace in which the relevant measurements are informationally complete. Doing so can result in reconstruction artifacts that originate in the state-space truncation and may result in inaccurate estimators for the unknown true state. Instead, one should perform the reconstruction on a larger subspace, with additional unsampled degrees of freedom, that is compatible with any prior information about a given unknown state. Such a more objective way of state estimation results in a much better tomographic quality of the reconstructed estimator.

\section{Acknowledgements}
This work is supported by the NUS Graduate School for Integrative Sciences and Engineering and the Centre for Quantum Technologies, which is a Research Centre of Excellence funded by Ministry of Education and National Research Foundation of Singapore, as well as the Czech Ministry of Education, Project LC06007, IGA Project PRF\underline{\,\,\,}2011\underline{\,\,\,}005, and the Czech Ministry of Industry and Trade, Project FR-TI1/364.

%%% just before \end{document}
%\cleardoublepage\newcommand{\twocolumn}{\relax}\newcommand{\onecolumn}{\relax}
%\input{LogLabs}\input{BibCheck}

\begin{thebibliography}{99}
\bibitem{remark1}
The word ``beam splitter'', used in this context, represents a class of possible apparatuses for splitting photon pulses, which includes conventional beam splitters, optical fibers, etc.
\end{thebibliography}
\end{document}